\documentclass[twocolumn,english,showpacs,preprintnumbers,groupedaddress,amsmath,amssymb,aps,prl]{revtex4}
\usepackage[T1]{fontenc}
\usepackage[latin9]{inputenc}
\usepackage{float}
\usepackage{textcomp}
\usepackage{amsmath}
\usepackage{graphicx}
\usepackage{amssymb}
\usepackage{esint}

\makeatletter
\@ifundefined{textcolor}{}
{%
 \definecolor{BLACK}{gray}{0}
 \definecolor{WHITE}{gray}{1}
 \definecolor{RED}{rgb}{1,0,0}
 \definecolor{GREEN}{rgb}{0,1,0}
 \definecolor{BLUE}{rgb}{0,0,1}
 \definecolor{CYAN}{cmyk}{1,0,0,0}
 \definecolor{MAGENTA}{cmyk}{0,1,0,0}
 \definecolor{YELLOW}{cmyk}{0,0,1,0}
 }

\makeatother

\usepackage{babel}

\makeatother

\usepackage{babel}

\begin{document}

\title{Optical precursors in the singular and weak dispersion limits:

 Reply to reply to comment}

\author{Bruno Macke}

\author{Bernard S\'{e}gard}

\email{bernard.segard@univ-lille-1.fr}

\affiliation{Laboratoire de Physique des Lasers, Atomes et Mol\'{e}cules , CNRS et
Universit\'{e} Lille 1, 59655 Villeneuve d'Ascq, France}

\begin{abstract}
The reply by Oughstun \emph{et al} . {[}J. Opt. Soc. Am. B \textbf{28}, 468-469 (2011){]} to our comment {[}J. Opt. Soc. Am. B \textbf{28}, 450-452 (2011){]} on their recently published paper {[}J. Opt. Soc. Am. B \textbf{27}, 1664-1670 (2010){]} is shown to make no response to the main points raised in our comment and even to bring additional confusion. The present reply provides the necessary clarifications.

 
\end{abstract}
\maketitle
In a recent comment \cite{ma11}, we pointed out inconsistencies in the paper by Oughstun \emph{et al.} on Sommerfeld and Brillouin precursors \cite{ou10}. Their reply \cite{ou11} makes no response to the main points raised in our comment. In addition it is written in such a manner that it may confuse and mislead the unfortunate reader. It thus appears necessary to dot the i's and cross the t's.

The first point concerns the analytical form of the Brillouin precursor. No need to look at the Oughstun's papers abundantly cited in \cite{ou10, ou11} for (possibly) finding the remarkably simple expression $ E_{B}(z,t')=\frac{b}{\omega_{c}}\mathrm{Ai}(-bt') $. As mentioned in our comment \cite{ma11}, this expression was established by Brillouin himself as far back as 1932 \cite{bri32}. The way by which we have retrieved this result shows that it will be a good approximation of the exact signal when the instantaneous frequency of the precursor is large (small) compared to the damping rate $\delta$ (the carrier frequency $\omega_{c}$). This condition is met in the singular dispersion limit, at least for the first oscillations of the precursor. Since $b\propto z^{-1/3}$ ($z$ propagation distance), the previous result \emph{analytically} shows that the amplitude of the Brillouin precursor also scales as  $z^{-1/3}$, a result that was obtained only \emph{numerically} in \cite{ou10}. The expression of $ E_{B}(z,t')$ also shows that the instantaneous period of the precursor is proportional to $1/b$. In particular, the first half-period $HP_{1}$ of the precursor, defined as the time interval between the first maximum and the first minimum of $ E_{B}(z,t')$, is equal to $2.22/b$. For the parameters used to obtain Fig.3 in \cite{ou10}, we get $HP_{1}=0.23$ ps in agreement with the exact result (see Fig.1 of our comment \cite{ma11}). On their side, Oughstun \emph{et al.} give in the equation (15) of \cite{ou10} an \emph{estimate} of the half-period that, in the same conditions, leads to a value smaller than the right one by about five orders of magnitude!

Contrary to the claim made in \cite{ou10}, we have shown in our comment \cite{ma11} that the precursors are catastrophically affected by the rise-time $T_{r}$  of the incident field even when the latter is considerably faster than the damping time $\tau =1/\delta$  of the medium. Our analysis was supported by numerical simulations made for rise times as fast as $\tau/500$  and  $\tau/100$. Without disputing the validity of our simulations, Oughstun \emph{et al.} claim in their reply  that our numerical results \textquotedblleft{}do not properly lie in the asymptotic regime of sufficiently large propagation distances\textquotedblright{} \cite{ou11}. In fact, the cases considered in our simulations are exactly those considered in their paper \cite{ou10}. In particular, as noticed both in the text and in the figure caption, the parameters of our Fig.3 are strictly identical to those of their Fig.3. However, as suggested by Oughstun \emph{et al.}, we have made additional simulations for larger propagation distances. These simulations show that the rise-time effects become more and more catastrophic when z increases. This result was obviously foreseeable. Indeed the effect of the rise-time, \emph{irrespective of the precise shape of the rise}, is to contract the spectrum of the incident field about the carrier frequency $\omega_{c}$. On the other hand, when $z$ increases, the medium becomes strongly absorbing on a broader and broader spectral region, keeping some transparency only at frequencies $\omega \gg \omega_{c}$ and $\omega \ll \omega_{c}$ . The precursors, associated with the transparency regions, then cannot be excited by the incident field.

Oughstun \emph{et al.} also invoke in their reply \cite{ou11} an eventual interaction between the Brillouin precursor and the so-called main signal. Such an interaction, if it exists, is extremely small in the conditions of our figure 3 for which the amplitude $A_{MS}$ of the main signal is only $\mathrm{e}^{-10}$, that is about $3700$ times smaller than the amplitude $A_{B}$ of the Brillouin precursor. The interaction becomes completely negligible when the propagation distance is increased in order to better \textquotedblleft{}lie in the asymptotic regime\textquotedblright{}. When the optical thickness at $\omega_{c}$ is equal to $100$ ($A_{MS}=\mathrm{e}^{-100}\approx 3.7\times10^{-44}$), we have determined the amplitude of the Brillouin precursor for $T_{r}=0$ and $T_{r}=\tau/500$ (error-function envelope signal) in the conditions of Fig.3 and Fig.5 in \cite{ou10}. These figures were intended to respectively illustrate the singular and weak dispersion limits. We find $A_{B}(0)\approx 7.9\times10^{-3}$ and $A_{B}(\tau/500)\approx 6.7\times10^{-10}$ in the former case;  $A_{B}(0)\approx 3.6\times10^{-2}$ and $A_{B}(\tau/500)\approx 2.2\times10^{-8}$ in the latter one. These results confirm \emph{in both cases} the catastrophic effect of the rise time of the incident field even when it is 500 times smaller than the damping time $\tau=1/\delta$ of the medium.
	
	In summary, the analysis presented in our comment \cite{ma11} on the paper \cite{ou10} is entirely confirmed.

\end{document}